\documentclass[12pt]{article}
\usepackage{latexsym,multicol,graphicx,rotating}

\usepackage{isomath}
\usepackage{upgreek}
\usepackage{txfonts} 
\usepackage{upgreek}
\usepackage{titling}

\title{How the minuscule can contribute to the big picture: the neutron electric dipole moment project at TRIUMF}

\author{Ruediger \textsc{Picker}$^{1,2}$ for the TRIUMF Japanese-Canadian UCN Collaboration \\
\small{$^{1}$TRIUMF, 4004 Wesbrook Mall, Vancouver, BC V6T 2A3, Canada} \\
\small{$^{2}$Simon Fraser University, 8888 University Drive, Burnaby, BC V5A 1S6, Canada}}

\begin{document}
\begin{titlingpage}
    \maketitle
    \begin{abstract}
        A permanent electric dipole moment (EDM) of a fundamental particle violates both parity (P) and time (T) reversal symmetry and combined charge and parity (CP) reversal symmetry if the combined reversal of charge, parity \textit{and} time (CPT) is preserved. It is a very promising place to search for physics beyond the Standard Model.
Ultracold neutrons (UCN) are the ideal tool to study the neutron electric dipole moment since they can be observed for hundreds of seconds.
This article summarizes the current searches for the neutron EDM using UCN and introduces the project to measure the neutron electric dipole moment at TRIUMF using its unique accelerator driven spallation neutron and liquid helium UCN source.
The aim is to reach a sensitivity for the neutron EDM of around $10^{-27} \,e \cdot$cm.
    \end{abstract}
\end{titlingpage}

\section{Introduction}
Permanent electric dipole moments (EDMs) give rise to fundamental symmetry violations: simultaneous parity (P) and time reversal (T) violation and therefore also combined charge and parity (CP) violation if combined charge, parity and time reversal symmetry (CPT) holds.
The Standard Model (SM) background for these violations is very low, predicting a neutron EDM at the level of $10^{-32}$ \rm{to} $10^{-31} \,e \cdot$cm;
many extensions of the SM contribute an EDM just below the current best experimental limit of $d_{\rm n} < 3 \times 10^{-26} \,e \cdot$cm (90\% CL) measured with ultracold neutrons \cite{nEDM2015}.

Additionally, CP violation can be related to the matter-antimatter asymmetry of the universe,
one of the most intriguing puzzles in fundamental particle physics, as well as philosophy.
Since Ramsey and Purcell published the first neutron EDM measurement~\cite{ramsey57}, an immense improvement of six orders of magnitude has been made.
Several projects worldwide aim to improve the current nEDM limit by at least an order of magnitude, mostly using ultracold neutrons (UCN).

UCN are neutrons of such remarkably small kinetic energies ($ < 300$~neV) that they can be stored in containers made of suitable materials, such as stainless steel, diamond-like carbon (DLC)~\cite{DLCpaper}, deuterated plastics~\cite{dPSpaper} etc and observed for several hundreds of seconds. 
So far, EDM measurements with ultracold neutrons have been statistics limited, hence advances in EDM precision are immediately linked to new generation, stronger UCN sources.
This contribution will describe the UCN/nEDM projects with a special focus on the TRIUMF UCN/EDM effort.

\section{Neutron electric dipole searches using ultracold neutrons}

As mentioned above, the current best limit for the nEDM has been determined at Institute Laue-Langevin (ILL) in Grenoble, France, by the \textbf{RAL/SUSSEX/ILL collaboration} ~\cite{ILLEDMapp2014,nEDM2015}.
The apparatus was using the UCN turbine at the ILL research reactor: a very cold neutron beam is extracted vertically from the ILL cold neutron source located close to the reactor core.
UCN are produced by a Doppler-shifting device, the Steyerl turbine~\cite{ILLUCNsource}, whose receding blades downshift the neutrons to lower energies.
As many other experiments, it uses Ramsey's method of separated oscillating fields~\cite{ramseymethod1950}:
the neutrons are polarized on their way from the source to the EDM experiment and stored inside a material trap where homogeneous and (anti-)parallel magnetic ($B_0$) and electric fields ($E_0$) are present.
They are then submitted to two short intervals of phase-coherent oscillating fields ($B_1$).
The first pulse rotates the polarization vector by $\pi/2$ to be perpendicular to $B_0$ and allows free precession.
After a predetermined time $T$, the second pulse, which is in phase with the first, rotates the polarization vector by another $\pi/2$ to become parallel to $B_0$, but only if the oscillation frequency is exactly on resonance with the Larmor precession of the neutron spin.
If not, this results in a non-complete flip of the neutron polarization.
Due to the long period (up to around hundred second) between the two pulses, the neutrons accumulate a large additional phase angle generating a large increase in contrast compared to the Rabi method~\cite{Rabi}.
Placing a polarization analyzer (often a magnetized iron foil creating a large enough field to reflect one spin state of the neutron) between the EDM cell and the detector, the polarization of the stored neutrons after the Ramsey cycle can be determined and with this the resonance frequency $\nu_0$ of the spins in the EDM cell.
Comparing these resonance frequencies between parallel and anti-parallel electric and magnetic field orientations give rise to the EDM signal as
\begin{equation}
d_{\rm n} = - \frac{\delta\nu_0 h}{4  E_0},
\end{equation}
where $h$ is Planck's constant.
The statistical sensitivity reach of this technique is
\begin{equation}
\sigma (d_{\rm n}) = - \frac{\hbar}{2 \alpha T E_0 \sqrt{N}}. \label{Sensit}
\end{equation}
$\alpha$ denotes the visibility of the central Ramsey fringe, which is unity in an ideal experiment, but reduced via e.g. depolarization of the neutrons; 
$N$ stands for the number of neutrons counted in the polarization sensitive detectors.

One of the key improvements of the RAL/SUSSEX/ILL measurement over previous ones was the use of a comagnetometer.
Spin-polarized $^{199}$Hg were stored simultaneously to the neutrons in the trap.
Similarly to the neutrons, a rotating field pulse $B_1'$ caused the spins to be nearly perpendicular to $B_0$.
Their free precession was probed with an optical method:
the transmission of light from $^{204}$Hg discharge lamp is dependent on the $^{199}$Hg spin orientation in the plane perpendicular to $B_0$.
A photo detector measured the transmitted light varying with time as an exponentially decaying sinusoid.
The $^{199}$Hg Larmor frequency average over the storage period was determined by a fit to the data and used to compensate for magnetic fields drifts.
Since the $^{199}$Hg EDM is orders of magnitude smaller than that of the neutron, the electric field interaction with mercury is negligible.

The \textbf{PSI EDM experiment} collaboration acquired the apparatus mentioned above, installed it at their spallation-target-driven UCN source at Paul-Scherrer-Institut, Villigen, Switzerland and made significant improvements to various aspects of the experiment~\cite{PSIntoHg2014,PSIAxions2015,PSIfalseHgEDM2015,PSISSA2015,PSIspinecho2015}.
Combined with the increased UCN density inside the trap from the solid-deuterium UCN source~\cite{PSIsource2015}, the projected statistical sensitivity by the end of 2016 is around $1 \times 10^{-26} \,e \cdot$cm~\cite{PSIKirch}.

At the same time, a new experimental apparatus is under development~\cite{n2EDMarxiv}: a cubic multi layer magnetic shield with a shielding factor of $10^5$ shall provide a B field homogeneity of 1 pT/cm.
The EDM spectrometer will be made up of a double precession chamber stacked vertically allowing measurements of the two B and E field configurations simultaneously.

The \textbf{EDM group at Technische Universit\"at M\"unchen}, Garching, Germany, uses a similar approach  with a magnetically shielded room created by two mu metal layers for low frequency and DC shielding of external magnetic fields and an aluminum layer in between for high frequency attenuation~\cite{fierlroom}.
Inside, an additional multilayer magnetic shielding houses a double chamber EDM experiment.
The experiment shall eventually be conducted at the UCN source of the Research Neutron Source Heinz Maier-Leibnitz (FRM II), Garching, which is currently under construction~\cite{FRM2source1,FRM2source2, FRM2source3}:
neutrons from the compact reactor core will be moderated in the surrounding heavy water and a solid hydrogen pre-moderator located inside a through-going beam tube at the (FRM II) generating a large flux of cold neutrons.
These are then converted to UCN via downscattering in a thin layer of solid deuterium and extracted via vacuum neutron guides.
With a projected UCN density of several thousand per cubic cm~\cite{fierlNovCim}, the statistical sensitivity reach will be around $5 \times 10^{-28} \,e \cdot$cm.

A \textbf{PNPI-ILL collaboration} (Petersburg Nuclear Physics Institute, Gatchina, Russia) has published the latest EDM limit of $5.5 \times 10^{-26} \,e \cdot$cm (90\% CL)~\cite{serEDM2015} using the PNPI double-chamber EDM spectrometer~\cite{PNPIspectr}. 
In the 1990's this spectrometer was the first to use two vertically stacked UCN chambers with a central high-voltage electrode to significantly suppress magnetic field fluctuation effects on the measurement:
common-mode magnetic noise affects both cells simultaneously and is therefore not contributing to the EDM signal to first order.
Eight magnetometers surrounded the EDM cells to estimate the overall field gradient.

The \textbf{Los Alamos National Laboratory (LANL) EDM project} builds on upgrades to the solid deuterium UCN source at LANSCE (Los Alamos Neutron Science Center), New Mexico~\cite{LANLUCNsource}.
Currently, the source delivers about 60 UCN/cm$^3$~\cite{ItoMainz}, which allows a statistical sensitivity of $5.6 \times 10^{-27} \,e \cdot$cm (90\% CL) to be reached in three years in a typical double cell EDM spectrometer.
Upgrades to the source that are currently implemented include improved guide coupling, optimized source and moderator design for an estimated increase of more than a factor of 2.5.
The project shall demonstrate the statistical feasibility of such a measurement and perform Ramsey cycle measurements.

A departure from the Ramsey method is pursued by the \textbf{SNS EDM project}~\cite{SNSEDM2014} to be located at the Spallation Neutron Source (SNS), Oakridge, Tennessee:
polarized $^3$He is used as comagnetometer and neutron spin analyzer at the same time.
SQUID magnetometers placed directly outside the EDM cells will sense the magnetic field variations created by the $^3$He spins all precessing in phase.
By measuring their precession frequency the average magnetic field inside the EDM cells is determined.
$^3$He is also a strong neutron absorber, but the absorption cross section $\sigma$ is very spin dependent $\sigma = \sigma_0 (1- a \cos{\vartheta})$, with $a \approx 1$ and $\vartheta$ is the angle between $^3$He and neutron spin: it is almost zero for parallel  alignment and largest for anti-parallel.
The reaction products of neutron capture $n + {\rm ^3He} \rightarrow {\rm ^3H} +  {\rm ^1H} + 0.764$~MeV scintillate in superfluid $^4$He, in which the experiment is conducted:
neutrons from the SNS cold source are converted into UCN by superfluid liquid helium directly inside the two EDM cells.
The beating frequency between the two spins modulates the scintillation light generation which can be measured using photo detectors.
A change of the beating frequency upon electric field reversal is a signal for the neutron electric dipole moment, since the $^3$He spin precession is not affected by the electric field.
An advantage of this method is the suppression of the sensitivity to magnetic field gradients and variations by an order of magnitude because $\frac{\gamma_{^3{\rm He}}-\gamma_{\rm n}}{\gamma_{\rm n}} \approx 0.1$, where $\gamma_{\rm n}$ and $\gamma_{^3{\rm He}}$ are the neutron and helium-3 gyromagnetic ratios, respectively.
Additionally, the Larmor precession frequencies of the two species can be synchronized by application of an additional oscillating magnetic field. 
In this critical dressing condition~\cite{spindressing}, the only relative spin rotation between He and n is due to a neutron EDM.
This further reduces the effect of the static magnetic fields on the measurement.
\begin{figure}[htb]
\includegraphics[width=\textwidth]{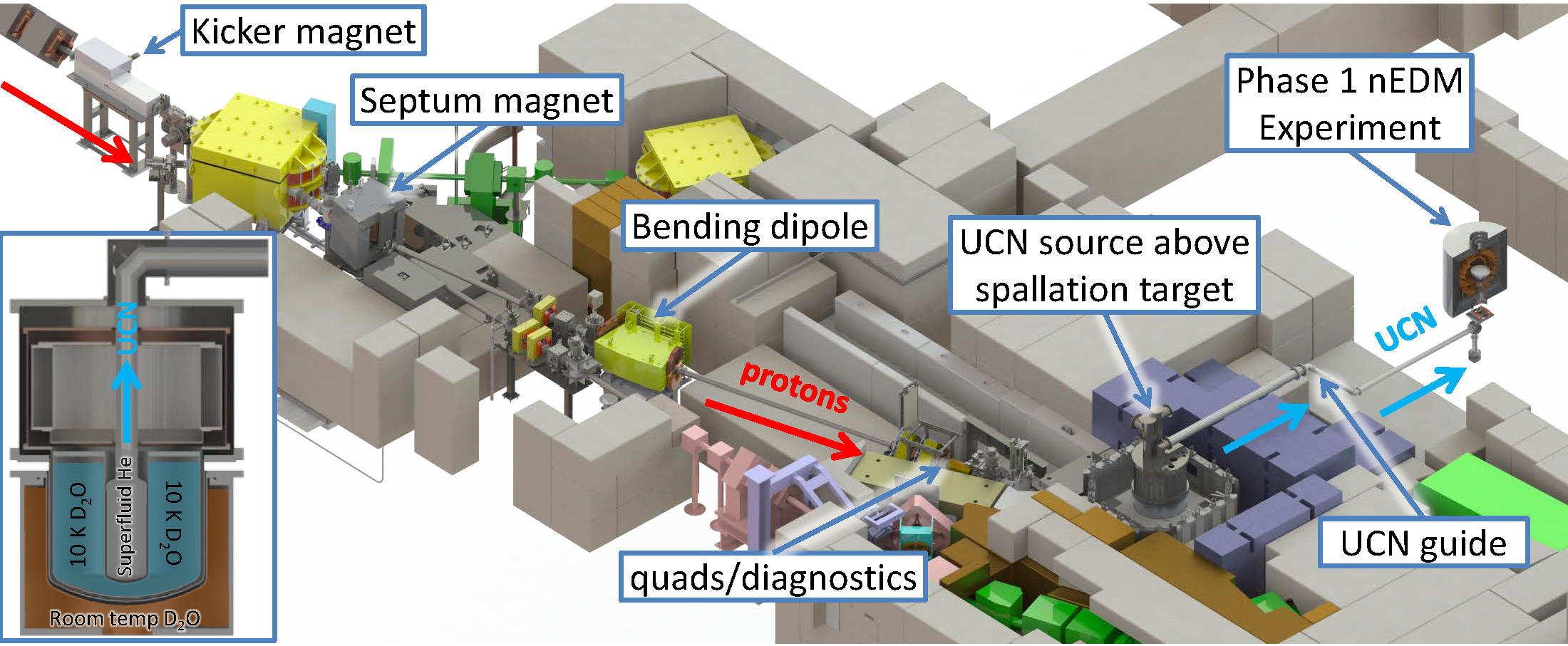}
\caption{Engineering drawing of UCN facility circa summer 2017. Protons from the TRIUMF cyclotron
are diverted by kicker, septum, and bender magnets onto the target. Neutrons liberated by spallation are
cooled in the UCN source apparatus. UCN are transported by guides to the Phase~I nEDM apparatus. Inset bottom left: schematic of the vertical UCN source cryostat.}
\label{TRIUMFPhase1}
\end{figure}

\section{The UCN source and EDM project at TRIUMF}
At TRIUMF, Vancouver, Canada a spallation-driven superfluid helium source is being built to feed a room temperature double cell Ramsey EDM apparatus.
The UCN source and EDM project are carried out in two phases each.

Buckets of 480-MeV protons from the TRIUMF cyclotron are diverted into the dedicated UCN beamline 1U by a fast kicker and a Lambertson septum. 
The kicker is able to ramp up to nominal current and down to zero in about 50~$ \upmu$s;
this is faster than the  $\approx$100~$\upmu$s gap in the proton beam from the cyclotron happening every ms.
At flattop it provides a kick of 14~mrad upwards so that the protons enter the upper (non-shielded) beam tube of the UCN septum magnet which creates a 9$^\circ$ deflection to the left.
The lower septum beam tube is surrounded by an iron yoke, shielding it from the magnet's field, allowing "unkicked" proton bunches to continues down beamline 1A serving the Centre for Molecular and Materials Science (CMMS) at TRIUMF.
This scheme allows variable extraction of 0 to 40~$\upmu$A into 1U out of the 120~$\upmu$A total current of the beamline, corresponding to a maximum of every third bucket being kicked into beamline 1U.
Further downstream, an additional dipole bender adds 7$^\circ$ more beam deflection creating sufficient space for the tantalum-clad tungsten spallation target, which also acts as a beam dump.
The spallation neutrons are reflected and moderated to room temperature by solid lead and graphite blocks surrounding the target and a heavy water tank.
A large flux of cold neutrons around 9~\AA \,\, is created by further moderation in solid heavy water in Phase~I of the UCN source and liquid deuterium in Phase~II.
Conversion to the ultracold regime happens in superfluid liquid $^4$He at around 0.8~K where the cold neutrons are downscattered creating phonons and rotons in the liquid.
The UCN are transported out of the source by total reflection on the surfaces of UCN guides.

Construction and initial commissioning of the beamline at TRIUMF occurred in 2016. 
The Phase~I UCN source will be installed 2017 with first UCN production expected in that summer.
Fig.~\ref{TRIUMFPhase1} shows the facility utilizing the vertical UCN source cryostat from the Research Center for Nuclear Physics (RCNP), Osaka, Japan that has proven to provide 26~UCN/cm$^3$ at the exit of the source~\cite{RCNPsource} above a lead target from 1~$\upmu$A of 400~MeV protons.
The ultracold neutrons are extracted vertically in this source eliminating the need for windows that the UCN have to pass on their way to the experiment:
the liquid helium is confined by gravity.
Three stages are necessary for cooling the isotopically pure $^4$He;
4.2~K are reached inside a liquid helium bath, about 1.1~K in a natural helium volume that is being pumped on and about 0.8~K in a small $^3$He reservoir connected to a large roots pump.
A copper heat exchanger finally cools the isotopically pure $^4$He to about the same temperature.
At TRIUMF, slightly higher proton energy and optimized target and solid moderator design should increase the UCN density compared to RCNP.
If the cooling power of the cryostat will allow it, the proton current will be increased beyond 1~$\upmu$A enabling a further boost in UCN production.
\begin{figure}[tb]
\begin{center}
\includegraphics[width=0.5\textwidth]{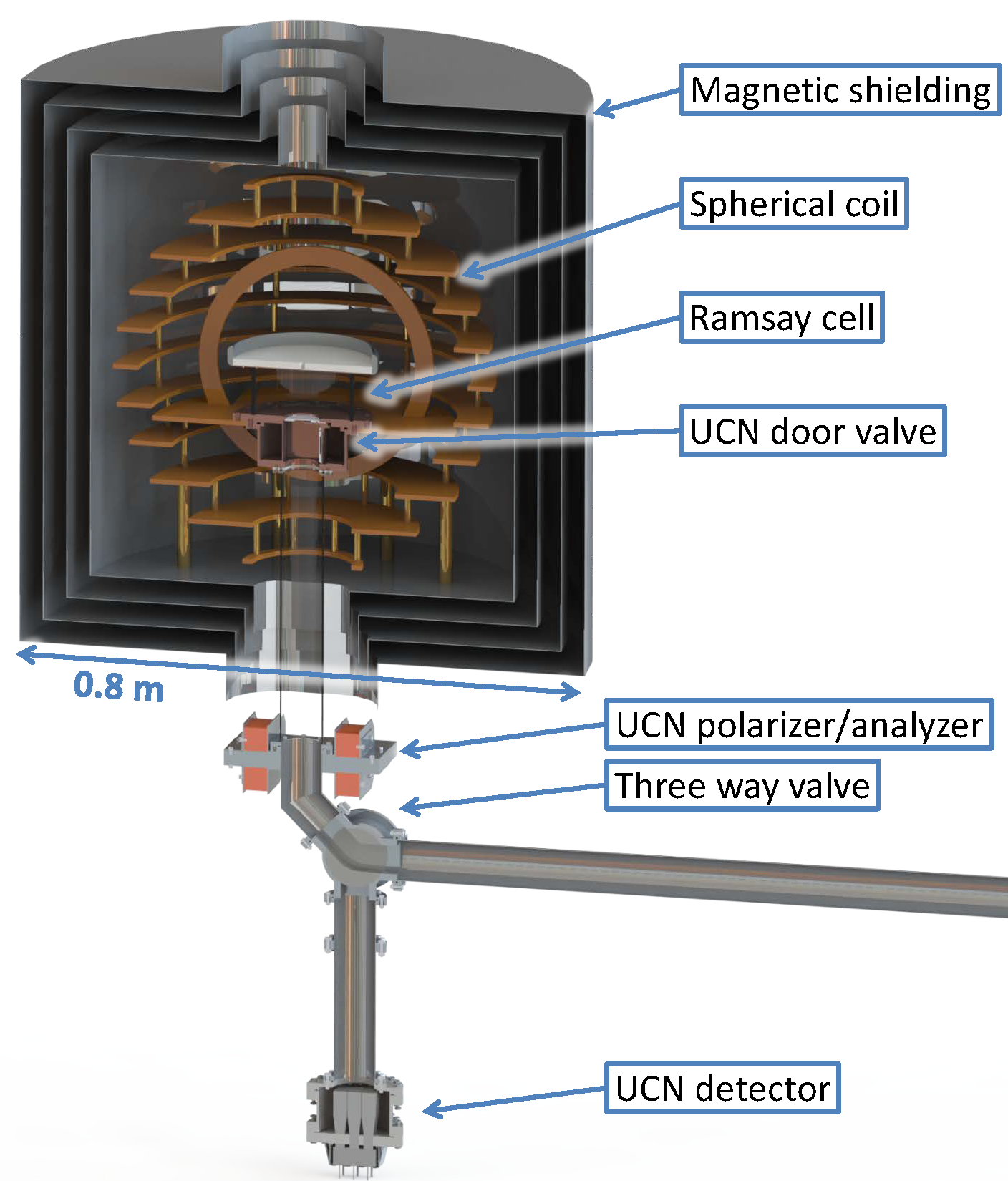}
\caption{Cut-away view of the EDM Ramsey apparatus: neutrons enter from the guide on the left, are diverted into the Ramsey cell via a three way valve and polarized on the way. After the Ramsay cycle the polarizer acts as analyzer and the three way valve diverts the UCN into the detector below.}
\label{Phase1EDM}
\end{center}
\end{figure}

Fig.~\ref{Phase1EDM} shows a cut-view of the Phase~I EDM apparatus from RCNP~\cite{RCNPEDM2} that will be installed 2017.
A four layer mu-metal shield surrounds the coils necessary for neutron Ramsey cycles:
a spherical coil to produce the homogeneous static magnetic field $B_0$ and a pair of Helmholtz coils for the oscillating $B_1$ field. 
A quartz cell with a door shutter serves as UCN storage cell.
Below the magnetic shielding a magnetized iron foil polarizes the neutrons on the way to the cell and is used as analyzer after the Ramsey cycle has finished.
At TRIUMF, the UCN will be detected in Li glass scintillators connected to photomultipliers~\cite{UWLidet}.
This apparatus will mainly be used as a technology development platform for the Phase~II experiment:
neutron handling, active magnetic shielding, high-voltage and comagnetometer hardware will be designed and optimized using this platform.

\begin{figure}[htb]
\includegraphics[width=\textwidth]{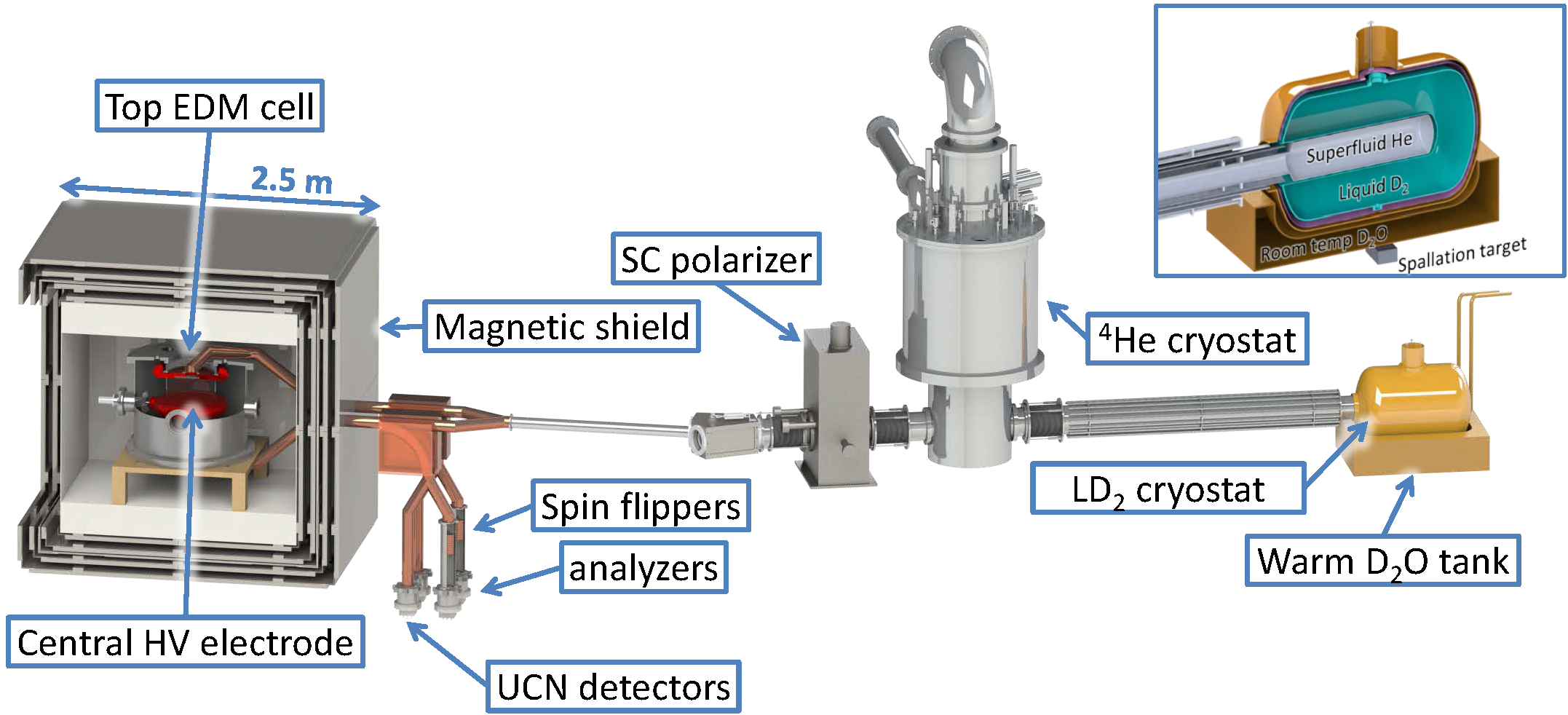}
\caption{3D view of the Phase~II UCN source and EDM apparatus. Inset top right: schematic of the liquid deuterium and horizontal UCN source upgrade.}
\label{Phase2}
\end{figure}

The Phase II nEDM experiment at TRIUMF builds on several significant UCN source upgrades and a completely new nEDM apparatus, sketched in Fig.~\ref{Phase2}.
Scheduled for 2019, the heavy water cryostat will be replaced by a liquid deuterium (LD2) volume that surrounds the UCN production volume and creates a large flux of cold neutrons.
LD2 has been proven one of the best cold neutron moderators at various facilities~\cite{LD2-1,LD2-2,LD2-3,LD2-4,lD2PSI}.
In the TRIUMF geometry compared to the heavy water, the UCN yield in the superfluid helium is expected to be more than a factor of five higher using LD2.
This factor takes into account the larger heat input into the superfluid, causing higher helium temperatures and hence a lower UCN storage time due to upscattering in the liquid~\cite{GolubHeUpsc} and assumes a UCN production volume made of beryllium walls.
The cryostat cooling the isotopically pure $^4$He will be placed a few meters away from the spallation target to avoid excessive activation.
Extraction of the UCN out of the liquid helium will  horizontal so that a larger phase space of UCN can be utilized in experiments.
Fig.~\ref{Phase2} shows a setup using aluminum foils to contain the helium and separate the source from the vacuum UCN guides.
Transmission through the foils will be facilitated by a 3.5~T superconducting polarizing magnet that allows only one spin state of the neutrons to pass and additionally accelerates these UCN through the foils reducing transmission losses.

A multilayer magnetically shielded room will create an ideal low-field magnetic environment, which the neutrons enter from the side:
four high permeability metal layers with a large DC and low-frequency shielding factor for magnetic fields and gradients~\cite{JBPassiveShielding} and an aluminum layer serving to shield high frequencies take the approach of \cite{fierlroom} one step further.
A self shielded $B_0$ coil~\cite{BidSelfShielded} will minimize the interaction of the internal field with external disturbances such as temperature changes of the mu metal.
The EDM spectrometer will be a double cell configuration with a central high-voltage electrode separated by two cylindrical insulator rings from two grounded electrodes creating opposite alignments of B and E field in the two cells.

Magnetic field fluctuations will be monitored in situ and online by a $^{199}$Hg comagnetometer similar to \cite{ILLEDMapp2014}, but using the Faraday rotation method as is used for mercury EDM experiments~\cite{HgEDM1,HgEDM2}.
In addition, the application of a dual comagnetometer of $^{199}$Hg and $^{129}$Xe is being pursued.
$^{129}$Xe has the advantage of having a much lower neutron absorption cross section:
21~barn compared to 2150~barn for $^{199}$Hg allowing a pressure of a few $10^{-3}$~mbar.
The relevant transition of $^{129}$Xe is a two photon transmission at $254.2$~nm.
Absorption at this wavelength is modulated by the xenon Larmor precession in the magnetic field and therefore also the resulting IR emission at 823~nm and 895~nm (lifetime of the states $\approx 2.5$~ns).
Monitoring infrared emission by a photo detector outside of the cell allows determination of the $^{129}$Xe Larmor precession frequency and therefore the average magnetic field in the cell.
Combining the two comagnetometer measurements allows to determine both the homogeneous component and the first order gradient of the magnetic field.

Polarization analysis of the neutrons after the completion of each Ramsey cycle will be done by a simultaneous spin analyzer system for each cell as used by the PSI experiment~\cite{PSISSA2015}:
above the detector(s), the UCN guide splits, one arm has a magnetized iron foil to only allow one neutron polarization to pass to one detector, the other arm has a spin flipper and a magnetized foil to allow the other polarization to pass to another detector.
Neutrons in the wrong arm are reflected and get another chance to be counted in the correct detector thereby increasing the statistical significance of the measurement.
\begin{figure} [htb]
  \begin{minipage}[b]{.5\textwidth} 
    \includegraphics[width=0.95\textwidth]{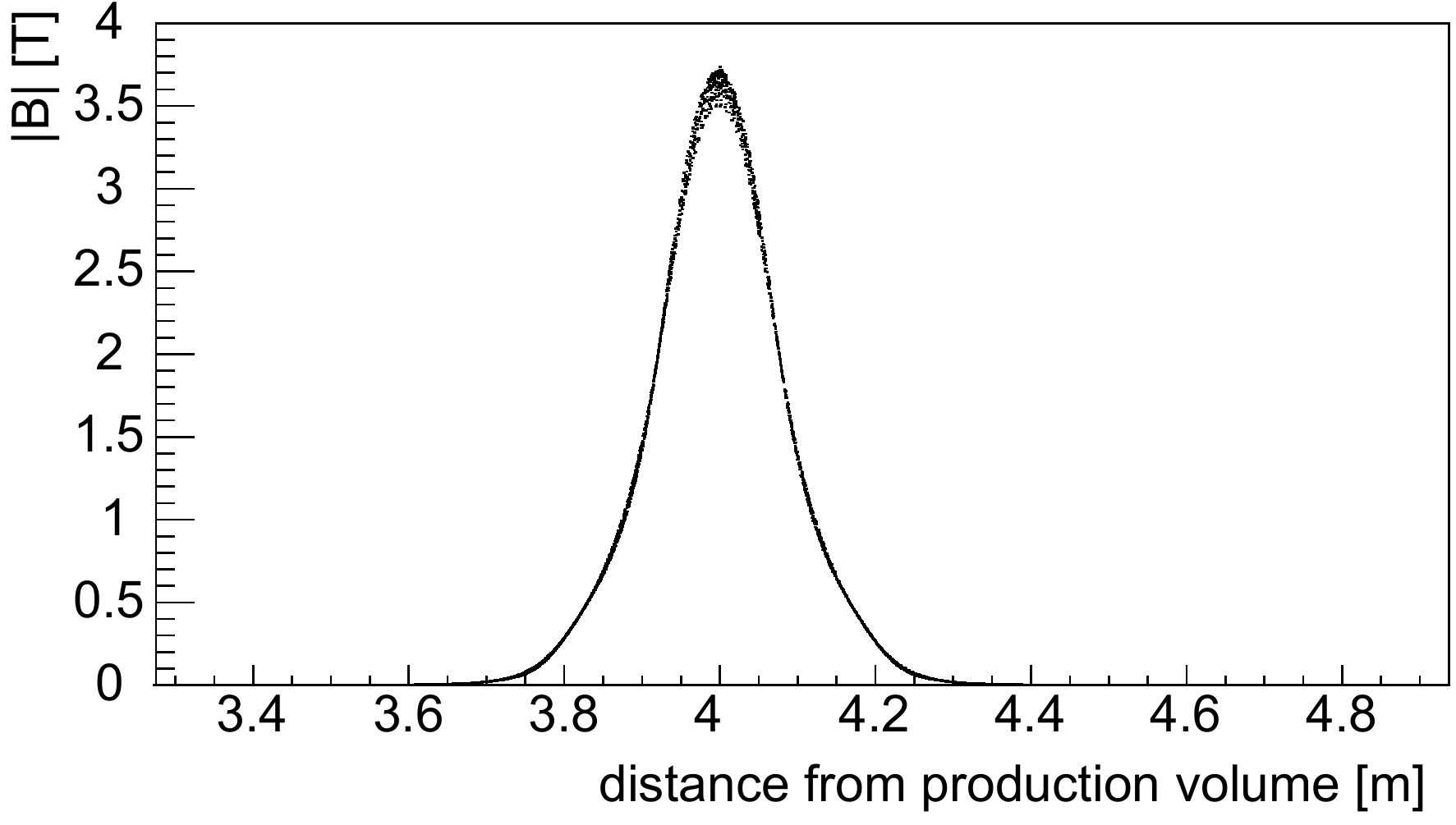} 
  \end{minipage} 
   \begin{minipage}[b]{.5\textwidth} 
    \includegraphics[width= 1\textwidth]{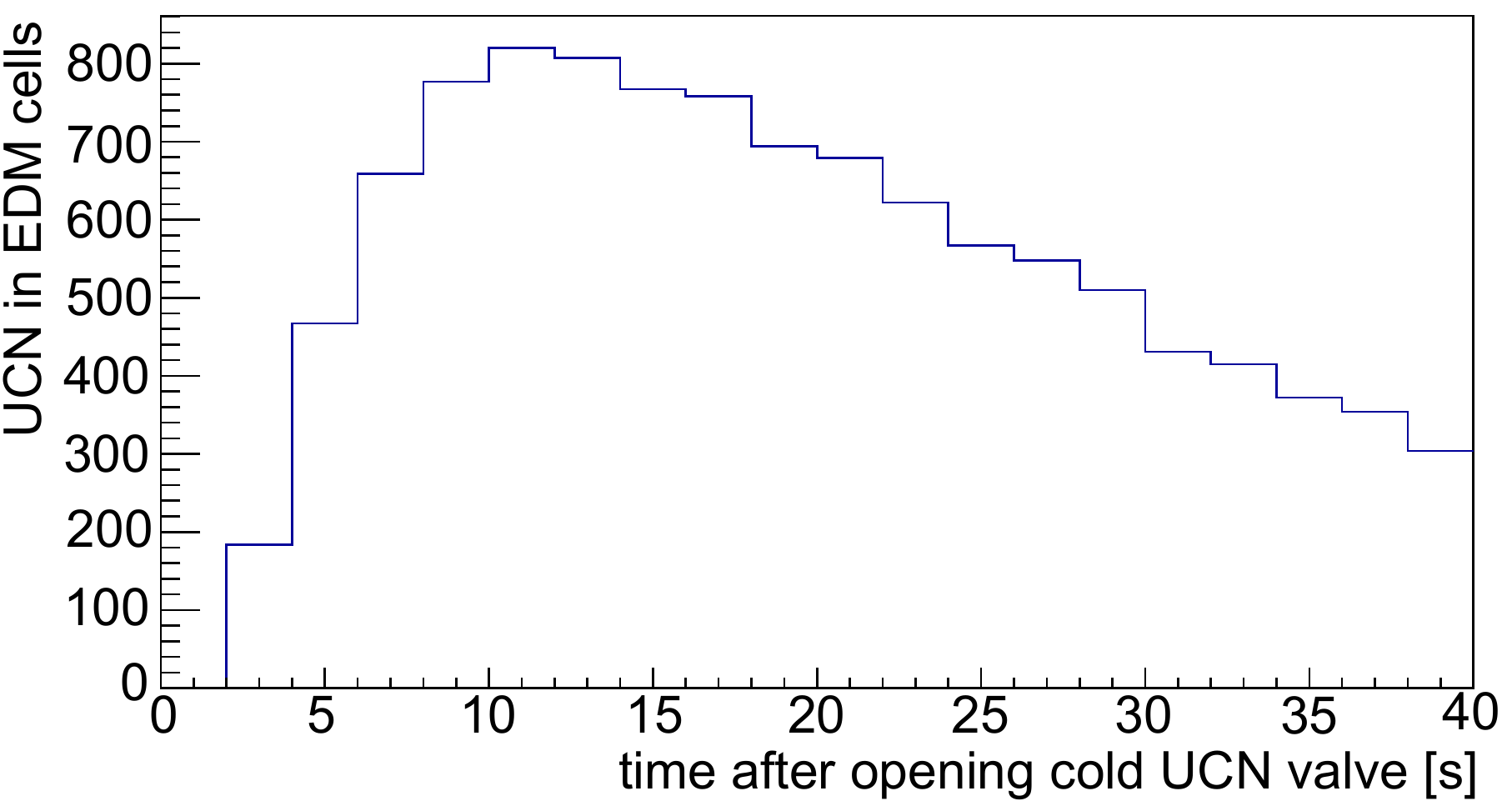} 
  \end{minipage}
  \caption{(Left) -  Magnetic field of the superconducting polarizer as used in the Monte Carlo simulation. (Right) - Time evolution of the number of UCN in the EDM cells after opening the He-II UCN valve. } 
 \label{FieldTime}
  \end{figure}

The following describes a comprehensive estimation of the statistical sensitivity that is possible in Phase~II of the EDM experiment at TRIUMF, which is planned to be commissioned at the turn of the decade.
It comprises several steps, from the proton beam to the spallation target, the moderators, UCN production, transport and storage in the EDM cells.
The calculation is based on the following assumptions, models and simulations:
The Phase~II UCN source is installed on beamline BL1U at TRIUMF, which is providing 40~$\upmu$A of 480~MeV protons to the tungsten spallation target.
Detailed MCNP6 simulations were conducted to determine the cold neutron flux in the UCN production volume.
From this flux, the UCN production rate is calculated as described in \cite{korobkina2002production} to be $\approx 2.3 \times 10^{7}$~UCN/s.
The resulting UCN spectrum in the UCN source is assumed to be the tail of a Maxwellian spectrum from 0 to 300~neV.
The $^3$He content in the $^4$He source volume is negligible in terms of UCN absorption, the $^4$He temperature is 0.8~K, resulting in an upscattering lifetime of a little less than 600~s.
During the 60~s irradiation period of the spallation target, the UCN are stored upstream of a UCN valve located inside the helium cryostat as shown in Fig.~\ref{Phase2}.
The valve is opened right after the proton beams stops.
UCN source guides are coated with DLC, as are all the other guide components towards the EDM experiment, except inside the $^4$He cryostat, which is lined with natural nickel (Fermi potential of 245~neV). 
DLC coated electrodes and deuterated polystyrene insulators comprise the EDM cells.
The wall loss lifetime of UCN upstream of the valve during proton beam irradiation is around  140~s for UCN of less than 245~neV.
This value is possible, since upscattering is largely suppressed at 0.8~K.
The superconducting polarizer as it exists at RCNP is at nominal current with a maximum field of around 3.5~T. Fig.~\ref{FieldTime} (left) shows a scatter plot of the absolute value of the magnetic field along the UCN guide as seen by UCN in the Monte Carlo simulation PENTrack~\cite{PENTrack}.
\begin{figure}[htb]
\begin{center}
\includegraphics[width=\textwidth]{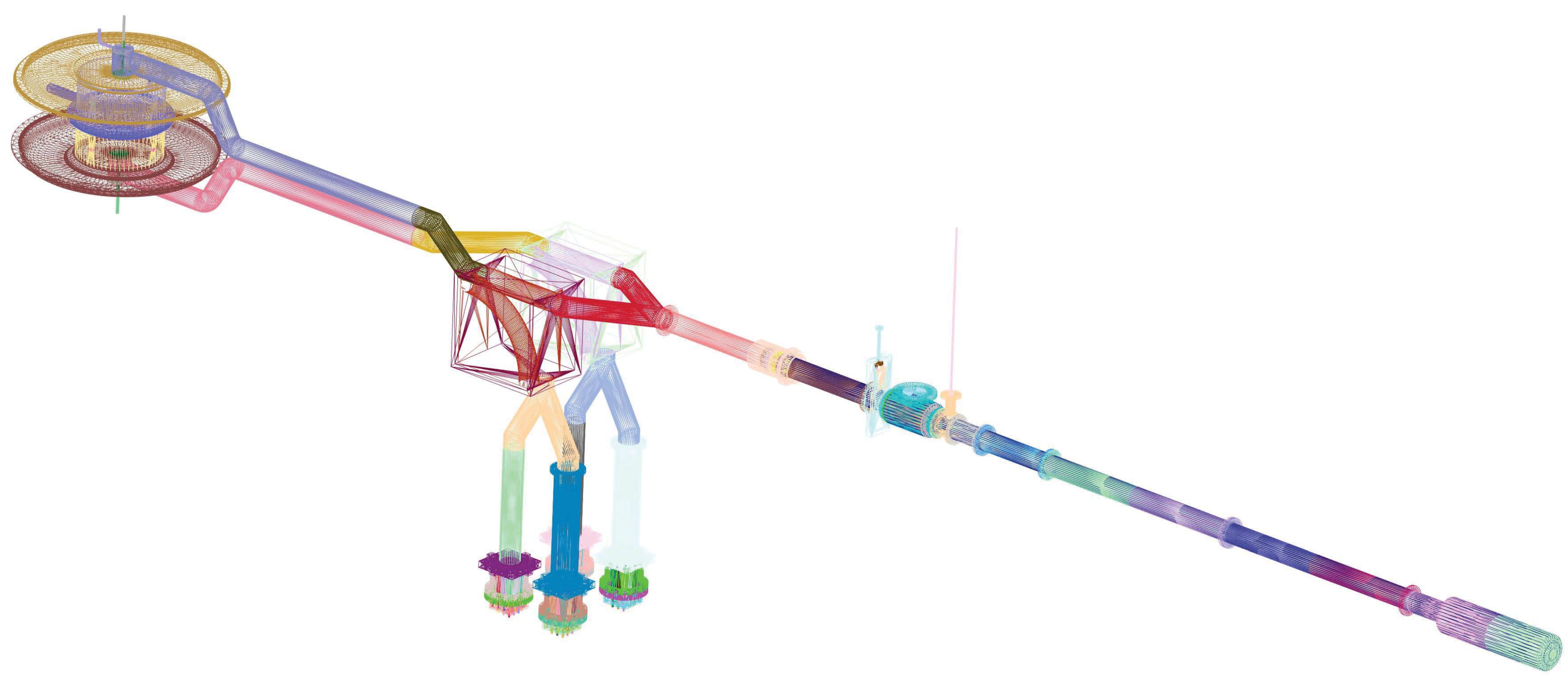}
\caption{\label{fig:STLTriangles} Geometry definition for the UCN Monte Carlo simulation PENTrack via STL files.}
\end{center}
\end{figure}
The magnetic field is calculated by a detailed Opera 3D model of the polarizer at RCNP which includes the return yoke and is imported into the UCN simulation as tabulated values on a 3D grid.
PENTrack uses a tricubic interpolation routine for calculating the magnetic field between the grid points.
 The geometry for the UCN simulation is created in Solidworks and exported into STL files (STL: Standard Tessellation Language) after creating triangular meshes. Fig.~\ref{fig:STLTriangles} shows the mesh of the simulation model. 
Two EDM cells with a diameter of 36~cm and a height of 15~cm, vertically stacked form the central experimental region.
An ideal straight UCN guide topology is assumed with a total distance of around 7.5~m from the UCN source to the EDM cells.
The inner guide cross sections are 8.5~cm.
In PENTrack, the interaction with material boundaries and bulk material is handled via determining UCN track intersections with the STL mesh triangles and invoking the relevant behaviors: specular and diffuse reflection (Lambert or Micro roughness model~\cite{Steyerl1972,HeuleMR}), transmission via Snell's law or diffuse (Lambert or Micro roughness model), absorption or upscattering at a material boundary or in the bulk of the material.
Gravitational interaction on the neutron and magnetic interaction on the neutron spin is used for neutron tracking.
The fields in the simulation are smooth enough so that the adiabatic spin transport condition is valid and a magnetic scalar potential can be used: $\mathbf U_{\rm B} = - \mu_{\rm n} |\mathbf B|$, where $\mu_{\rm n}$ is the magnetic moment of the neutron.
        Three aluminum foils separating the liquid helium in the UCN source from the vacuum UCN guides are implemented in the simulation, the two outer ones having a thickness of 100~$\upmu$m, the central one 10~$\upmu$m.
         The complex Fermi potential values for the materials in the simulation are taken from measurements where available or calculated from neutron scattering lengths~\cite{NScatt}.        
 The filling time was determined to be a little more than 10~s as shown in Fig.~\ref{FieldTime} (right).
  A transport efficiency of around $4$\% was found using the PENTrack simulation as described above, resulting in a UCN density of 680~UCN/cm$^3$ or a total number of $2.1 \times 10^{7}$ UCN in both EDM cells after filling.
 The storage lifetime in the EDM cells are assumed to be around $\tau = 85$~s, the spin relaxation times $T_1 = 2000$~s and $T_2 = 1000$~s, the initial visibility $\alpha_0 = 0.95$ and the UCN storage period in the cells is set to $t_{\rm c}=100$~s.
The losses on the way to the detector have been determined via Monte Carlo simulation to be 10\%, the detector efficiency is assumed to be 90\%.
              
According to Eq.~\ref{Sensit} this results in a statistical sensitivity of $\sigma(d_{\rm n}) = 5.6 \times 10^{-26} \,e \cdot$cm per cycle.
One EDM cycle comprises eight fills of the EDM cells, four to determine the resonance frequency for each electric field orientation.
After around 100 days of beam time, a sensitivity of $\sigma(d_{\rm n}) = 1 \times 10^{-27}\,e \cdot$cm can be reached assuming 14 hours of stable magnetic field environment nightly.

During Phase~II, the UCN facility at TRIUMF shall be equipped with a second experimental port, creating a user facility that will be open to the worldwide UCN community.

\section{Conclusions}
The search for the electric dipole moment of the neutron is as important as ever:
this tiny quantity can help solve major physics and philosophical puzzles and change our big picture of the Universe.
Several promising efforts using ultracold neutrons are ongoing worldwide to push the current limit down by another order of magnitude into the $d_{\rm n} < 10^{-27} \,e \cdot$cm region within the next five to ten years.









\end{document}